\documentclass[prl,aps,graphics,a4paper,10pt,twocolumn,nofootinbib]{revtex4-1}

\usepackage[margin=25mm]{geometry}
\usepackage[table,xcdraw]{xcolor}
\usepackage{amsmath}
\usepackage{braket}
\usepackage{hyperref}

\usepackage{tikz}
\usepackage{pgfplots}
\usepackage{tikzscale}
\usepackage{ mathrsfs }

\newcommand{\bqa}{\begin{eqnarray}}
\newcommand{\eqa}{\end{eqnarray}}

\usepackage{comment}
\usepackage{placeins}

\newcommand{\ens}{E} 

\newcommand{\be}{\begin{equation}}
\newcommand{\ee}{\end{equation} }
\DeclareMathOperator{\tr}{Tr}

\newcommand{\kb}[2]{|#1\rangle\langle#2|}

\begin{document}

\title{Quantum Simulation of the Dissipative Anderson Model}

\author{Max Hunter-Gordon}
\author{Zsolt Szab\'o}
\author{Robert A. Nyman}
\author{Florian Mintert}

\affiliation{Physics Department, Blackett Laboratory, Imperial College London, Prince Consort Road, SW7 2AZ, United Kingdom}

\begin{abstract}
The interplay of Anderson localisation and decoherence results in intricate dynamics but is notoriously difficult to simulate on classical computers. We develop the framework for a quantum simulation of such an open quantum system making use of time-varying randomised gradients, and show that even an implementation with limited experimental resources results in accurate simulations.
\end{abstract} 

\maketitle

Decoherence due to environmental noise can fundamentally change the character of dynamics in quantum systems, from the decay of otherwise stable states up to the emergence of classical dynamics. 
Our understanding of decoherence processes is limited by the difficulties that we are facing trying to simulate the dynamics of open quantum systems.
The advent of quantum simulators offers the possibility to simulate coherent~\cite{QS} or dissipative~\cite{Barreiro:2011ve} dynamics of physical models  that are prohibitively expensive to simulate on existing classical computers.

A particularly striking interference phenomenon is the transition of perfectly delocalised Bloch waves, in periodic potentials, to exponentially localised eigenstates in the presence of weak disorder that breaks a system's periodicity.
This Anderson localisation~\cite{Anderson} typically degrades in the presence of decoherence such that an initially localised wave-packet can spread over the entire system. Since generically coherent superpositions of wave-packets with large spatial separation decay faster than coherent superpositions of wave-packets with small spatial separation, such a dephasing Anderson system can feature an intricate interplay of coherent dynamics on short spatial scales but incoherent dynamics on larger scales.
In one-dimensional systems, this is expected to result in the growth and subsequent decay of interference peaks that exist neither in the perfectly coherent nor in the dephased system~\cite{2018arXiv181010588R}.

For higher-dimensional or even interacting systems, there is extremely limited knowledge about the interplay between Anderson localisation and decoherence.
While the analysis of Anderson localisation in two or three dimensions is a formidable computational challenge on its own~\cite{3D_Anderson},
the inclusion of decoherence effects is likely to keep exceeding our computational capabilities for the foreseeable future, and calls for a quantum simulation.

With Anderson localisation observed in several highly controllable systems including (classical) acoustic and light waves \cite{WEAVER,PhysRevLett.100.013906,Lightwaves}, Bose-Einstein condensates in speckle potentials and optical lattices \cite{Billy,Roati2008AndersonCondensate}, the experimental prerequisites for a quantum simulation of the dephasing Anderson model seem to be available.
We therefore develop the framework for the realisation of such a quantum simulation.
In order to remain specific, we will discuss this in reference to atoms trapped in an optical lattice,
but most of the concepts derived here are platform agnostic and apply equally well to {\it e.g.} networks of superconducting qubits \cite{kjaergaard2019superconducting} or photonic circuits~\cite{Cuevas:2019vn}.

The Anderson Hamiltonian 
describes the dynamics of a single particle through an approximately periodic potential.
As long as the disorder is sufficiently weak, one can define states $\ket{x}$ that correspond to the particle being localised in the potential minimum at position $x$.
With the disorder resulting in energy shifts of the different potential minima, the Anderson Hamiltonian~\cite{Anderson} reads 
\be
H=\sum_{x} \epsilon_{x} \kb{x}{x}+\tau \sum_{\langle x,y\rangle}\left( \kb{y}{x} + \kb{x}{y}\right)\ ,
\label{eq:H}
\ee
where $\epsilon_{x}$ are the random energy shifts, $\tau$ is the rate of tunnelling processes between different lattice sites, and the summation $\sum_{\langle x,y\rangle}$ is performed over nearest neighbour sites.
The dimensionality of the underlying model enters Eq.~\eqref{eq:H} only in terms of the connectivity, {\it i.e.} the number of nearest neighbour sites. 
The Anderson Hamiltonian arises naturally for bosonic atoms in optical lattices with an additional speckle potential.
Typically these systems are described by the Bose-Hubbard Hamiltonian
\be
H_H=\sum_{x} \epsilon_{x}a_x^\dagger a_x+\tau \sum_{\langle x,y\rangle}\left( a_y^\dagger a_x + a_x^\dagger a_y\right)+H_I\ ,\nonumber\ee
with annihilation(creation) operators $a_x^{(\dagger)}$ for a particle at site $x$, satisfying bosonic commutation relations, and an onsite interaction $H_I$.
In the single-particle limit this Hubbard Hamiltonian reduces to the above Anderson Hamiltonian with the states $\ket{x}$ defined as $\ket{x}=a_x^\dagger \ket{0}$.

Dephasing will be taken into account in terms of a Lindbladian $\mathscr{L}$ of the form
\begin{equation}
\mathscr{L}(\kb{x}{y}) = - \gamma f(x,y)\kb{x}{y},
\label{eq:L}
\end{equation}
where $\gamma$ defines the time-scale of the dephasing processes and $f(x,y)$ is a function that captures the difference in dephasing rate for different coherent superpositions.
The condition $f(x,x)=0$ reflects that $\mathscr{L}$ includes only loss of phase coherence but no diffusive processes,
and a large(small) value of $f(x,y)$ implies that the phase coherence between the states $\ket{x}$ and $\ket{y}$ decays particularly fast(slow).

\begin{figure}
\centering
\includegraphics[width=0.5\textwidth]{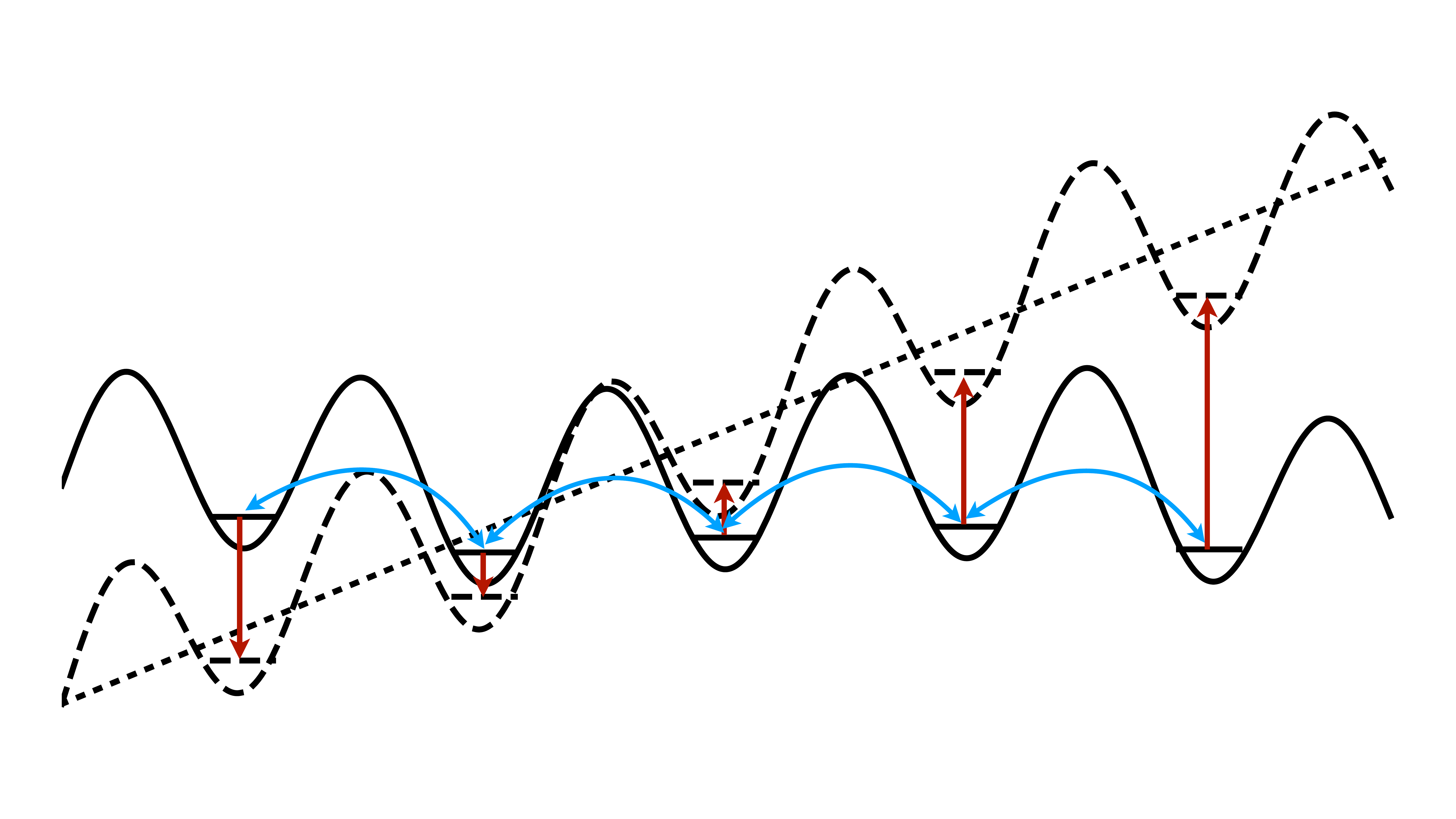}
\caption{Lattice potential with a small amount of spatial disorder (black, solid line).
Particles can occupy the ground state of any local potential well and tunnel between the corresponding sites (blue, curved arrows).
A linear potential, {\it i.e.} tilt (black, dotted line) together with the lattice potential creates a tilted lattice potential (black dashed line).
The energy shift resulting from this tilt (red, directed arrows) induces a site-dependent phase evolution.
Since the difference in phase evolution for two sites increases with distance, an ensemble average over different tilt strength results in the desired dephasing process.}
\label{fig:tilt}
\end{figure}

A key element of optical lattice experiments is the close-to-perfect elimination of decoherence~\cite{RevModPhys.80.885}.
It is thus necessary to introduce a mechanism resulting in decoherence, and to do so in a fashion that permits tuning $\gamma$, the strength of decoherence.
This can be achieved in terms of an energy gradient (or equivalently an acceleration) \cite{PhysRevLett.100.080404} as depicted in Fig.\ref{fig:tilt}, that is made to fluctuate.
Such a tilt, in one-dimension, is described in terms of the Hamiltonian
\be
H_{T} = \alpha \sum_{x} x \kb{x}{x}\ ,
\label{eq:tilt}
\ee
with the parameter $\alpha$ characterising the strength of the tilt.

For the sake of simplicity, let us consider for the moment a very strong optical lattice such that tunnelling is negligible.
Since the onsite energy part $\sum_{x} \epsilon_{x} \kb{x}{x}$ of the Anderson Hamiltonian commutes with $H_T$, we can thus discuss the dynamics induced by the tilt alone.
Any matrix element of the system state $\rho(t)$ after propagation for time $t$ then reads
\be
\bra{x}\rho(t)\ket{y}=\bra{x}\rho(0)\ket{y}\ e^{i\alpha(x-y)t}\ .
\ee
This is still perfectly coherent dynamics, but an ensemble average over different tilt strengths $\alpha$ will result in the desired dephasing process.
To this end, one can take the tilt $\alpha$ to be a random variable with Gaussian distribution centred around $\alpha=0$ with width $\sigma$.
For the ensemble-averaged state $\varrho$ this yields
\be
\bra{x}\varrho(t)\ket{y}=\bra{x}\varrho(0)\ket{y}\ \exp\left(-\frac{\sigma^2 t^2}{2}(x-y)^2\right)\ .\nonumber
\label{eq:gauss}
\ee
Thus, attenuation of phase coherence with a decay depending on the distance between $\ket{x}$ and $\ket{y}$ is obtained.
Phase coherence, however, does not decay exponentially in time, as expected for a Lindbladian, but there is a Gaussian time-dependence.
This issue can be overcome in terms of a time-dependent width $\sigma$ \cite{PhysRevA.92.032111},
or by adopting a stroboscopic perspective in which the system is probed only at integer multiples of some time constant $T$.
If the ensemble average resulting in the attenuation of phase coherence of Eq.~\eqref{eq:gauss} is performed independently in each interval of duration $T$, then observation after $n$ periods, {\it i.e.} after the duration $t=nT$, will yield an attenuation of phase coherence by a factor $\exp\left(-n\sigma^2 T^2/2\ (x-y)^2\right)$. This is
consistent with the Lindbladian defined above in Eq.~\eqref{eq:L} with the choice $f(x,y)=(x-y)^2$ and $\gamma=\sigma^2 T/2$.

In principle, this dephasing mechanism is enough to implement the desired quantum simulation based on a Trotter decomposition in terms of time windows of tilt dynamics, with suppressed tunnelling, alternating with time windows of dynamics induced by the Anderson Hamiltonian with finite tunnelling but no tilt.
In practice, however, it is not even necessary to modulate the depth of the lattice, but the phase-averaging effect described above can also be realised very well in the presence of finite tunnelling, simply because the Trotter decomposition asserts that the dynamics induced by a sequence of two alternating Hamiltonians coincides with the dynamics induced by the sum of the two Hamiltonians.
It is thus possible to realise the present quantum simulation in a digital fashion, {\it i.e.} alternating between Hamiltonians $H$ and $H_T$, and in an analogue fashion with no alternation between tilt and tunnelling.

Both realisations simulate the desired dynamics in the limit of a perfect ensemble average, infinitely fast switching of tilts ({\it i.e.} $T\to 0$ with $\gamma=\sigma^2 T/2$ constant) and in the case of the digital realisation, infinitely short Trotter steps.
As we will show in the following, however, even simulations with averages over rather small ensembles, reasonably long time-windows $T$ and few Trotter steps give an excellent account of the desired dynamics.

\begin{figure*}
\includegraphics[width=\textwidth]{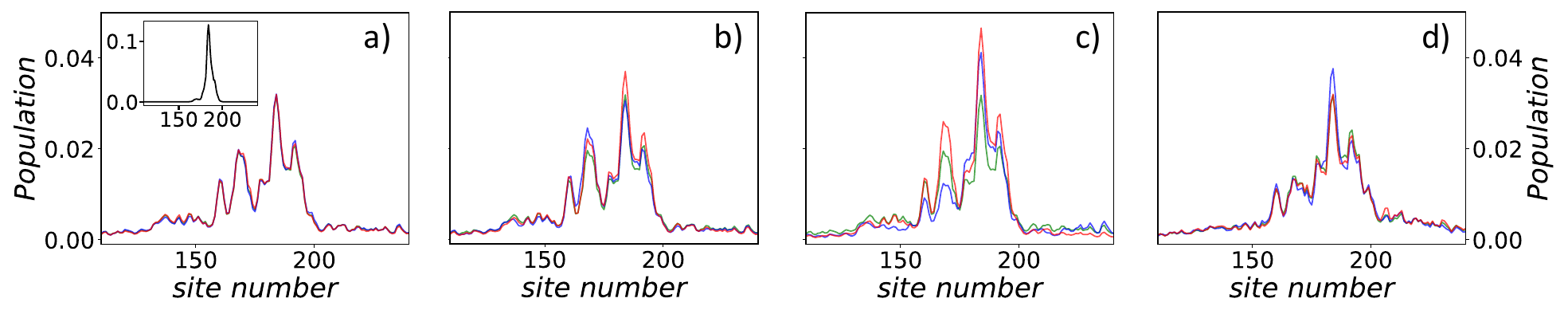}

\centering
\caption{Site populations after the central peak of the initial state (inset of a)) has decayed to $1/4$ of its original height corresponding to exact dynamics (green), digital (blue) and analogue (red) quantum simulations. Cases a) to c) correspond to $\gamma =10^{-4}\tau$ and total time $t = 653 \tau$; d) correspond to $\gamma =10^{-3}\tau$ and total time $t = 120 \tau$.
Due to decoherence an initially localized state widens, and develops fine interference structures that will decay at longer times only.
The different cases correspond to different targeted population infidelities with values of roughly $0.003$ in a), $0.01$ in b) and d), and $0.03$ in c).}
\label{fig:gamma}
\end{figure*}

In all the subsequent discussion we will consider a linear chain with $400$ sites and onsite energies $\epsilon_x$ drawn from a uniform distribution in the interval $[- \tau/5, \tau/5]$;
the system is initialised in the ground state of the Anderson Hamiltonian $H$ (given in Eq.~\eqref{eq:H}).
Fig.~\ref{fig:gamma} shows the comparison of site occupations obtained from dynamics induced by the exact Lindbladian (shown in green) and from an ensemble average over Hamiltonian dynamics including tilts corresponding to a digital(blue) and analogue(red) quantum simulation.
The decoherence rate $\gamma$ is chosen to be substantially smaller than the tunnelling rate $\tau$ such that the system displays aspects of coherent dynamics. 

Fig.~\ref{fig:gamma} a) depicts a very accurate quantum simulation where the population dynamics are modelled nearly perfectly. There are minimal deviations between the different methods, corresponding to $N=80$ steps in the trotterisation and an ensemble of $\ens=400$ different tilt configurations for the digital method, and $N=100$, $\ens=500$ for the analogue method.

A realization with
$N=20$, $\ens=100$ in the digital case and $N=30$, $\ens=300$ in the analogue case, as depicted in Fig.~\ref{fig:gamma} b),
reproduces the locations and qualitative shape of all interference structures very well. The reduced effort in the quantum simulation results in only slight misestimation of the exact height of the interference pattern.

Fig.~\ref{fig:gamma} c) depicts a case of a very rough quantum simulation,
with $N = 5$, $\ens = 25$ for the digital method, and $N = 15$ and $\ens = 150$ for the analogue method.
This strong reduction in experimental effort results in quantitative 
deviations between the quantum simulations and the exact dynamics, but all qualitative features like position and structure of interference peaks are reproduced much better than one might have expected at the given inaccuracy.

Since the digital method achieves simulation of the dephasing dynamics without an extra Trotter-approximation, it manages to achieve a given fidelity with fewer Trotter steps than the analogue method.
This, however, comes at the price of necessity to modulate the amplitude of the optical lattice, and it may depend on the details of an explicit implementation whether the digital or the analogue method seems preferable.

In order to verify that the numerical accuracy observed in Fig.~\ref{fig:gamma} is not specific to a given realisation of disorder it is helpful to define an infidelity for statistical analysis.
The accuracy of the populations depicted in Fig.~\ref{fig:gamma} is most appropriately characterized in terms of the population infidelity $I_p=(\sum_x\bra{x}\delta\ket{x}^2)^\frac{1}{2}$,
where $\delta=\varrho_e-\varrho_a$ is the difference of quantum state $\varrho_e$ resulting from the exact Lindbladian dynamics and quantum state $\varrho_a$ resulting from the averaged Hamiltonian dynamics.
For quantum states that are diagonal in the $\ket{x}$ basis, this is equivalent to the state infidelity $I_s=(\tr\delta\delta^{\dagger})^\frac{1}{2}$,
which for general states also assesses how well off-diagonal elements are being reproduced.

The explicit infidelities for the examples in Fig.~\ref{fig:gamma} read:
\be
\begin{array}{|c|c|c|c|c|}\hline
&a)&b)&c)&d)\\\hline
I_p \mbox{(digital)} & 0.0036 & 0.013 &0.032 & 0.014 \\\hline
I_p \mbox{(analog)} & 0.0032& 0.014 &0.038 & 0.0057\\\hline
I_s \mbox{(digital)}& 0.049 & 0.098 & 0.20 & 0.095\\\hline
I_s \mbox{(analog)}& 0.042 & 0.075 &0.15 & 0.055\\\hline
\end{array}\nonumber
\ee
These figures shall help to gauge to what accuracy of quantum simulation the infidelities in the following analysis correspond.

\FloatBarrier
\begin{figure}[h]
\centering
\includegraphics[width=0.4\textwidth]{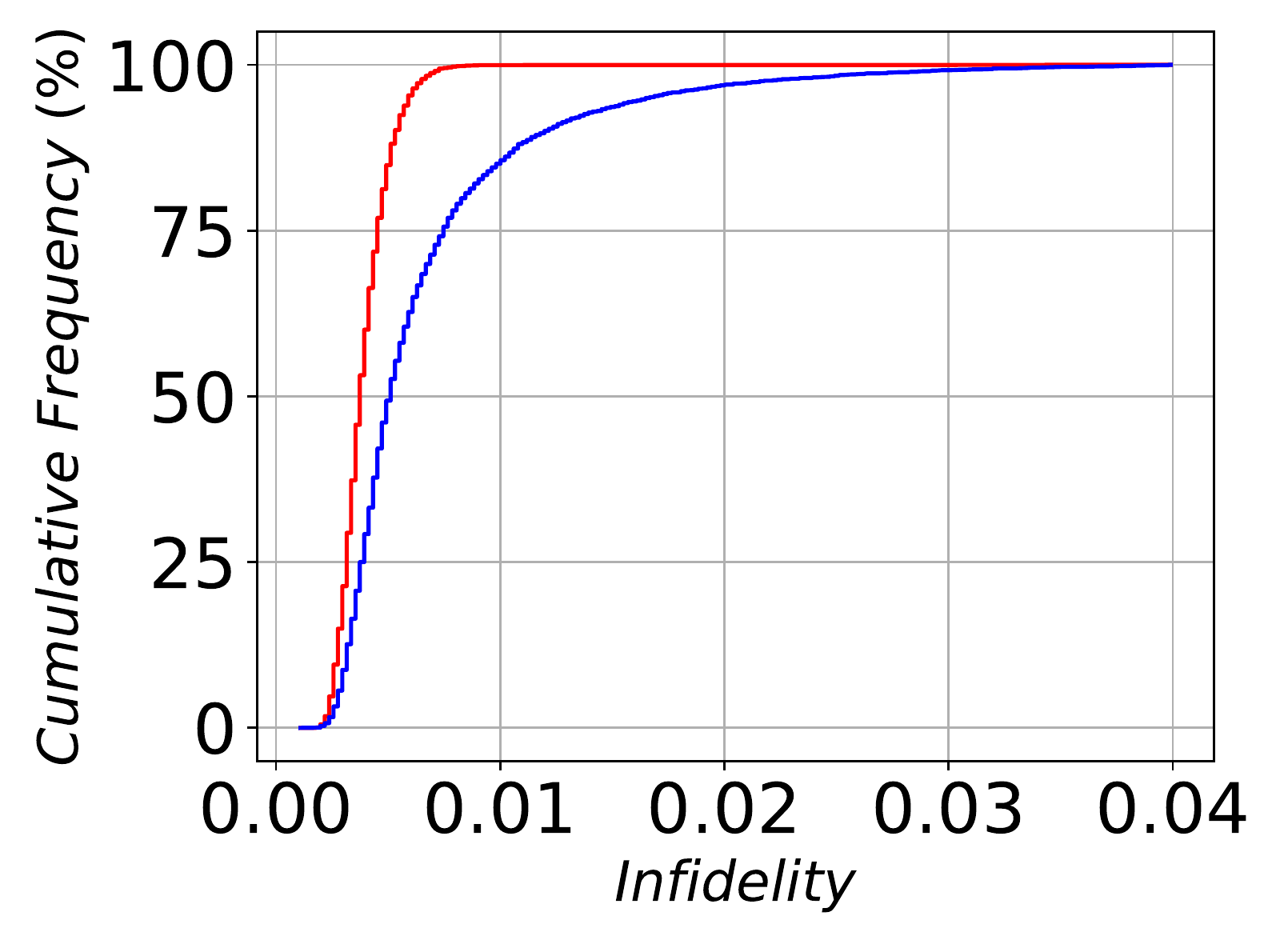}
\caption{Distribution for population infidelities of quantum simulation with $\ens=400$ and $N=80$ in the digital case (red) and with $\ens=500$ and $N=100$ in the analogue case (blue).
In both cases the $10,000$ different realisations of disorder result in typical infidelities between $0.002$ and $0.01$.
The distributions have median values of around $0.0039$ and $0.0052$ for the digital and analogue methods respectively.
Nearly all the digital quantum simulations yield infidelities below $0.01$, however a significant proportion of the analogue simulations result in an infidelity larger than this value.}
\label{fig:fidelities}
\end{figure}

The distribution of population infidelities obtained with $10,000$ different realisations are depicted in Fig.~\ref{fig:fidelities}. Results obtained with digital(analogue) quantum simulations are depicted in red(blue). As one can see, the use of $E=400(500)$ ensemble members and $N=80(100)$ Trotter steps results in typical population infidelities between $0.002$ and $0.01$. In the digital case, the distribution is narrow with a rapidly decaying tail. Typical infidelities in the analogue case are not significantly higher than those of the digital case yet the tail is more pronounced; that is, with the analogue method, there is some risk to encounter a specific disorder realisation that results in an atypically inaccurate quantum simulation.
This is not necessarily a surprise, because the equivalence of dephasing and averages over tilts with a finite number of Trotter steps is given rigorously only in the digital case, and it is rather astonishing that the analogue method, that comes with substantially reduced experimental complexity, is nearly as good as the digital method for the vast majority of disorder realisations.

The comparison between the exact Lindbladian dynamics and averaged Hamiltonian dynamics, presented here, is necessarily restricted to a one-dimensional system that allows for sufficiently efficient numerical simulation.
Whereas an explicit implementation of the present protocol would provide the proof of principle for the quantum simulation of the interplay between Anderson localisation and decoherence,
any realisation with a higher-dimensional or an interacting system would help us to explore physics that becomes prohibitively difficult to simulate by classical means.
Natural questions to be explored with such a platform could include signatures of the mobility edge in the presence of dephasing or interaction induced stabilisation of structures that would decay in the non-interacting system.

The decoherence model considered here can readily be generalised to any dependence on spatial separation in terms of the distribution for the average over different tilts.
Controlled decoherence can evidently also be realised with many different mechanisms, such as photon scattering \cite{PhysRevLett.93.073904}.
The quantum simulation envisioned here aims at reproducing the behaviour consistent with a given Lindbladian, but one may similarly also consider a system-environment interaction as an underlying model that could be realised
with a second species of trapped atoms serving as environment.
Controlling the inter-species interaction and/or temperature of the environmental species would then allow tuning of the decoherence time, and could be used to explore the transition from Markovian to non-Markovian dynamics~\cite{RevModPhys.88.021002}.

{\it Acknowledgements} We are indebted to stimulating discussions with Yannic Rath and Kiran Koshla.
This work has received funding from the European Unions Horizon 2020 research and innovation programme under grant agreement No. 820392 (PhoQuS).

\bibliography{mybib}
\end{document}